\documentclass[article,aps,pra,reprint,twocolumn,superscriptaddress,10pt]{revtex4-1}
\usepackage{amsmath}
\usepackage{mathrsfs}
\usepackage{amsfonts}
\usepackage{amssymb}
\usepackage{graphicx}
\usepackage[colorlinks,linkcolor=blue,anchorcolor=blue,urlcolor=blue,citecolor=blue]{hyperref}
\usepackage{eufrak}
\usepackage{multirow}
\usepackage{mathrsfs}
\usepackage{textcomp}
\usepackage{appendix}
\usepackage{makecell}
\usepackage{multirow}

\begin{document}

\title{Deterministic generation of multi-qubit entangled states among distant parties using indefinite causal order}

\author{Wen-Qiang Liu}\email{Contact author: wqliu@stdu.edu.cn}
\affiliation{Department of Mathematics and Physics, Shijiazhuang Tiedao University, Shijiazhuang 050043, China}
\author{Hai-Rui Wei}\email{Contact author: hrwei@ustb.edu.cn}
\affiliation{School of Mathematics and Physics, University of Science and Technology Beijing, Beijing 100083, China}

\date{\today }

\begin{abstract}
Quantum entanglement plays an irreplaceable role in various remote quantum information processing tasks. Here we present protocols for generating deterministic and heralded $N$-qubit entangled states across multiple network nodes.
By utilizing a pre-shared maximally entangled state and single-qubit operations within an indefinite causal order framework, the multi-qubit entangled state between distant parties can be generated deterministically.
The complex entangled state measurements and multiple pre-shared entangled states, are essential in conventional entanglement swapping technique, but are not required in our approach. This greatly reduces the complexity of the quantum circuit and makes it more experimentally feasible.
Furthermore, we develop optical architectures to implement these protocols by encoding qubits in polarization degree of freedom. The results indicate that our protocols significantly improve the efficiency of long-distance entanglement generation and provide a practical framework for establishing large-scale quantum networks.
\end{abstract}


\maketitle

\section{Introduction}  \label{sec1}

Quantum entanglement \cite{horodecki2009quantum} over long-distance quantum network nodes is a key resource for implementing distributed quantum computing  \cite{jiang2007distributed,oh2023distributed,liu2024nonlocal}, quantum secure communication \cite{long2002theoretically,xu2020secure,sheng2022one,zhou2022one,zhou2023device,zhao2024quantum}, quantum nonlocal correlations \cite{chaturvedi2024extending,villegas2024nonlocality,lobo2024certifying}, and quantum metrology \cite{yin2020experimental,zhao2021field,malia2022distributed}.
Nowadays, many works have been proposed for generating entangled states \cite{wang201818,chen2024heralded,cao2024photonic}.
However, these theoretical and experimental works primarily focus on qubit-based local systems, while long-distance entanglement states remains relatively scarce.
Entanglement swapping is a well-known technique to create long-distance entanglement between two remote particles \cite{pan1998experimental,de2005long,su2016quantum}. By utilizing two pre-shared entangled pairs and an appropriate entanglement measurement (such as, Bell state measurement), entangled states between two distant qubits can be generated.

Over the past two decades, entanglement swapping has been experimentally demonstrated across various physical platforms, including all-photonic systems \cite{pan1998experimental,liu2022all}, atoms \cite{chou2007functional}, trapped ions \cite{riebe2008deterministic}, quantum dots \cite{zopf2019entanglement}, and superconducting circuits \cite{ning2019deterministic}.
The primary challenges in entanglement swapping lie in the efficient generation of entangled photon pairs and the implementation of Bell state measurement (BSM).
First, the generation efficiency of entangled photon pairs using spontaneous parametric down-conversion (SPDC) is inherently low \cite{couteau2018spontaneous}.
Second, the probabilistic character of BSM is unavoidable in linear optical systems without additional resources or assistance \cite{lutkenhaus1999bell,calsamiglia2001maximum,bayerbach2023bell}.
To completely distinguish all Bell states, strategies such as utilizing additional degrees of freedom (DoFs) \cite{walborn2003hyperentanglement,barbieri2007complete,williams2017superdense,zhou2022deterministic}, or additional entanglement, or nonlinear media \cite{sheng2010complete,li2016complete,li2019resource} are usually adopted.
However, these approaches are challenged by inefficiency and impracticality.

Besides the quantum swapping technique, a quantum bus assisted by auxiliary photons provides an alternative approach for creating remote multi-qubit entanglement \cite{callus2021cumulative,li2018quantum,li2016heralded,sangouard2011quantum,munro2012quantum,wu2022quantum,xie2023heralded,zhou2023parallel,li2024heralded}.
In this approach, a hybrid entanglement between a stationary qubit and a photon qubit is established at each node, followed by photon-photon interactions at an intermediate node \cite{callus2021cumulative} or photon-spin interactions at another node \cite{li2018quantum}. After appropriate measurements are made on auxiliary photons, the system will automatically collapse into an entangled state.
Unfortunately, such multi-qubit entanglement is degraded by the channel noise and photon loss, because the numerous single photons or entangled photon pairs need to travel in long-distance optical fibers or free-space channels \cite{li2016heralded,sangouard2011quantum,munro2012quantum,wu2022quantum}.
In 2019, Piparo \emph{et al.} \cite{lo2019quantum} proposed quantum multiplexing to improve the efficiency of the long-distance entanglement generation.
Later, some intriguing protocols utilizing an entangled photon pair \cite{xie2023heralded} or a single photon as a quantum bus \cite{zhou2023parallel,li2024heralded} have been proposed to improve the efficiency of multiple entangled pairs.
Long-distance quantum entanglement has been experimentally demonstrated in various systems \cite{bernien2013heralded,van2022entangling,krutyanskiy2023entanglement}, but the limited photon transmission rate restricts the efficiency of the multi-qubit long-distance entanglement generation.

In this paper, we present alternative protocols for deterministically generating an $N$-qubit entangled state among $N$ spatially separated parties over long distances within an indefinite causal order (ICO) framework. ICO, a fascinating resource, shows significant advantages over fixed causal order approaches in various quantum information processing tasks \cite{wei2019experimental,nie2022experimental,yin2023experimental,liu2024experimentally,liu2025quantum}.
It is known that ICO is realized through quantum switch, and quantum switch has been demonstrated both theoretically \cite{chiribella2013quantum,chiribella2021indefinite,kechrimparis2024enhancing} and experimentally \cite{procopio2015experimental,goswami2018indefinite,guo2020experimental,stromberg2023demonstration,rozema2024experimental}. Our protocol begins with a pre-shared $N$-qubit maximally entangled state as the control qubits and one single particle held by each party as the target qubits.
Then, each party employs the control qubits to manipulate the order of single-qubit gates on their respective target qubits. Finally, by measuring the control qubits and adjusting an angle of the single-qubit rotation gates, the system can collapse into a maximally entangled state (MES), or a partially entangled state (PES), or a separable state (SS).
In entanglement swapping techniques, which require many pre-shared entangled pairs and complex joint measurements \cite{pan1998experimental,de2005long,su2016quantum,liu2022all,chou2007functional,riebe2008deterministic,zopf2019entanglement,ning2019deterministic}, our protocols rely solely on a pre-shared entanglement, the superposition of simple single-qubit gate orders, and straightforward single-qubit measurements. The approach eliminates the need for multiple entangled states and complex entangled state measurements, thereby reducing both resource requirements and measurement complexity.  These advantages make our protocols more practical for the implementation of remote entanglement generation.
Additionally, we develop an optical architecture to implement the protocols by utilizing the polarization and path DoFs of single photons. The efficiency of these protocols for $N=3$ and $N=4$ achieve exponential improvements.









%

\section{Entangled state generation between multiple long-distance qubits} \label{Sec2}

\subsection{Entangled state generation between two long-distance qubits}  \label{Sec2.1}

\begin{figure} 
\begin{center}
\includegraphics[width=8.2 cm,angle=0]{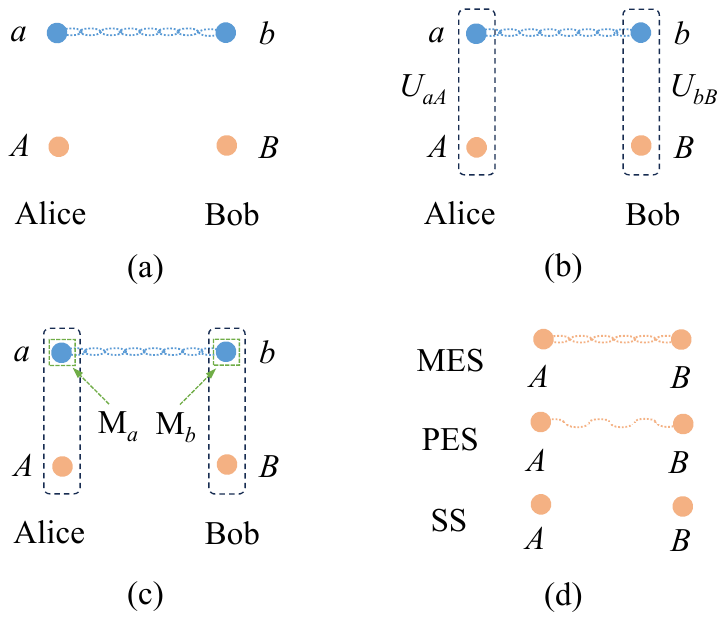}
\caption{Schematic diagram for generating a general two-qubit quantum state on two remote particles $A$ and $B$. (a) First, a maximally entangled state $|\varphi\rangle_{ab}=\frac{1}{\sqrt{2}}(|0\rangle_{a}|0\rangle_{b}+|1\rangle_{a}|1\rangle_{b})$ is shared with Alice and Bob, and Alice and Bob possess qubits $a$ and $A$, and $b$ and $B$, respectively. (b) Then, Alice performs a two-qubit unitary operation $U_{aA}$ on qubits $a$ and $A$, and Bob performs a two-qubit unitary operation $U_{bB}$ on qubits $b$ and $B$, respectively. (c)-(d) Lastly, the qubits $a$ and $b$ are measured (denoted by M$_a$ and M$_b$) in the basis $\{|+\rangle, |-\rangle\}$ and qubits $A$ and $B$ will automatically become a general two-qubit quantum state, i.e., a MES, or a PES, or a SS.}
\label{general}
\end{center}
\end{figure}

Suppose that Alice and Bob possess particle $A$ and $B$, respectively, and they are
initially prepared in the arbitrary states
\begin{eqnarray}            \label{eq1}
|\psi\rangle_A =\sqrt{\alpha_A} |0\rangle_A+\sqrt{1-\alpha_A} |1\rangle_A,
\end{eqnarray}
\begin{eqnarray}            \label{eq2}
|\psi\rangle_B =\sqrt{\alpha_B} |0\rangle_B+\sqrt{1-\alpha_B} |1\rangle_B.
\end{eqnarray}
Here the coefficients $\alpha_A, \alpha_B \in [0,1]$. Alice and Bob would like to change a given quantum state $|\psi\rangle_A \otimes |\psi\rangle_B$ to obtain a general two-qubit state, i.e., a MES, or a PES, or a SS.

Figure \ref{general} shows our protocol for creating a general state between two long-distance qubits $A$ and $B$ from a product state $|\psi\rangle_A \otimes |\psi\rangle_B$. Our protocol can be completed by the following steps.

First, as shown in Fig. \ref{general}(a), a maximally entangled state
\begin{eqnarray}            \label{eq3}
|\varphi\rangle_{ab}=\frac{1}{\sqrt{2}}(|0\rangle_{a}|0\rangle_{b}+|1\rangle_{a}|1\rangle_{b}),
\end{eqnarray}
is shared with Alice and Bob. The subscripts $a$ and $b$ denote that the qubits $a$ and $b$ are sent to Alice and Bob, respectively. Thus, Alice possesses qubits $a$ and $A$, and Bob possesses qubits $b$ and $B$.  The initial state of composite system composed of qubits $a$, $b$, $A$ and $B$ is given by
\begin{eqnarray}
\begin{split}             \label{eq4}
|\Phi_0\rangle = &\;|\varphi\rangle_{ab}\otimes|\psi\rangle_A\otimes|\psi\rangle_B\\
               =&\;\frac{1}{\sqrt{2}}(|0\rangle_a|0\rangle_b + |1\rangle_a|1\rangle_b)\otimes|\psi\rangle_A\otimes|\psi\rangle_B.
\end{split}
\end{eqnarray}

Second, as shown in Fig. \ref{general}(b), Alice (Bob) applies a local two-qubit gate $U_{aA}$ ($U_{bB}$) on qubits $a$ and $A$ ($b$ and $B$). After this, the state $|\Phi_0\rangle$ becomes
\begin{eqnarray}
\begin{split}             \label{eq5}
|\Phi_1\rangle =&\;\frac{1}{\sqrt{2}}[U_{aA}(|0\rangle_a\otimes |\psi\rangle_A)\otimes U_{bB}(|0\rangle_b \otimes|\psi\rangle_B) \\&
                                    + U_{aA}(|1\rangle_a\otimes |\psi\rangle_A)\otimes U_{bB}(|1\rangle_b\otimes |\psi\rangle_B)].
\end{split}
\end{eqnarray}

\begin{figure} 
\begin{center}
\includegraphics[width=8.2 cm,angle=0]{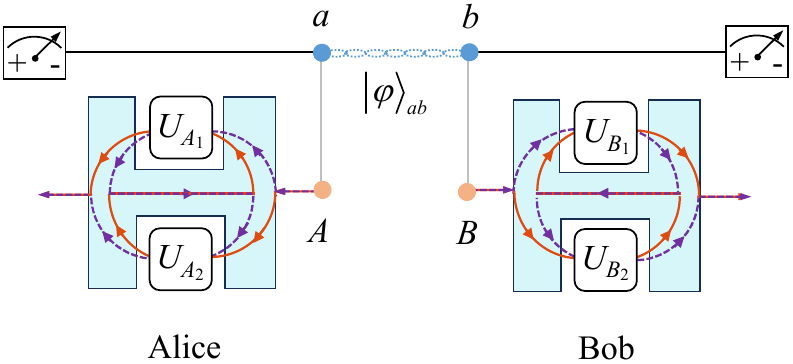}
\caption{Schematic diagram for generating a  heralded general two-qubit quantum state on two remote particles $A$ and $B$ via two quantum switches.
First, a maximally entangled state $|\varphi\rangle_{ab}$ is shared with Alice and Bob.
Qubits $a$ and $b$ act as the control qubits, which conditionally manipulate the operation order on target qubits $A$ and $B$ via their individual quantum switches, respectively.
When the control qubit $a$ ($b$) is in the state $|0\rangle_a$ ($|0\rangle_b$), the operation order of single-qubit gates is that $U_{A_1}$ ($U_{B_1}$) is before $U_{A_2}$ ($U_{B_2}$), i.e.,  $U_{A_2}U_{A_1}$ ($U_{B_2}U_{B_1}$), denoted by the purple circuit.
When the control qubit $a$ ($b$) is in the state $|1\rangle_a$ ($|1\rangle_b$), the operation order of single-qubit gates is that $U_{A_2}$ ($U_{B_2}$) is before $U_{A_1}$ ($U_{B_1}$), i.e.,  $U_{A_1}U_{A_2}$ ($U_{B_1}U_{B_2}$), denoted by the red circuit.
Lastly, the qubits $a$ and $b$ are measured in the basis $\{|\pm\rangle\}$, the target state of particles $AB$ will automatically become a MES, PES, or SS, by appropriately employing the single-qubit gates $U_{A_1}$, $U_{A_2}$, $U_{B_1}$, and $U_{B_2}$.}
\label{Switch}
\end{center}
\end{figure}

Finally, as shown in Figs. \ref{general}(c)-(d), the qubits $a$ and $b$ are measured in the basis $\{|\pm\rangle=\frac{1}{\sqrt{2}}(|0\rangle\pm|1\rangle)\}$. As a result, the state $|\Phi_1\rangle$ collapses to a two-qubit MES, or PES, or SS on $A$ and $B$.
In the protocol, the gates $U_{aA}$ and $U_{bB}$ as the core components determine the output states. These operations can be implemented using the technique of superposition of causal orders. To achieve this,  we employ quantum switches \cite{chiribella2013quantum,rozema2024experimental}, to cause the superposition of quantum gate orders. The implementation of this method is illustrated in Fig. \ref{Switch}.


As shown in Fig. \ref{Switch}, a pre-entangled state $|\varphi\rangle_{ab}$ connects two quantum switches held by Alice and Bob, respectively.
Each quantum switch uses $a$ ($b$) as the control qubit and $A$ ($B$) as the target qubit.
If the control qubit is in the state $|0\rangle_a$ ($|0\rangle_b$), the single-qubit gate $U_{A_1}$ ($U_{B_1}$) is applied before $U_{A_2}$ ($U_{B_2}$), resulting in the operation $U_{A_2}U_{A_1}$ ($U_{B_2}U_{B_1}$). This operation order is represented by the purple dashed circuit.
Conversely, if the control qubit is in the state $|1\rangle_a$ ($|1\rangle_b$), the single-qubit gate $U_{A_2}$ ($U_{B_2}$) is applied before $U_{A_1}$ ($U_{B_1}$), resulting in the operation $U_{A_1}U_{A_2}$ ($U_{B_1}U_{B_2}$). This operation order is represented by the red solid circuit.
Hence, two quantum switches transform $|\Phi_0\rangle$ into
\begin{eqnarray}
\begin{split}             \label{eq6}
|\Phi_1'\rangle =&\;\frac{1}{\sqrt{2}}[|0\rangle_a (U_{A_2}U_{A_1}|\psi\rangle_A)\otimes |0\rangle_b (U_{B_2}U_{B_1}|\psi\rangle_B) \\&
                                     + |1\rangle_a (U_{A_1}U_{A_2}|\psi\rangle_A)\otimes |1\rangle_b (U_{B_1}U_{B_2}|\psi\rangle_B)].
\end{split}
\end{eqnarray}
Based on Eq. \eqref{eq5} and Eq. \eqref{eq6}, it can be observed that the two-qubit gates $U_{aA}$ and $U_{bB}$ can be implemented by a coherently controlled quantum switch. These gates satisfy the following relationship:
\begin{eqnarray}
\begin{split}             \label{eq7}
U_{aA} =&\;|0\rangle_a\langle0|\otimes (U_{A_2}U_{A_1})+|1\rangle_a\langle1|\otimes (U_{A_1}U_{A_2}),
\end{split}
\end{eqnarray}
\begin{eqnarray}
\begin{split}             \label{eq8}
U_{bB} =&\;|0\rangle_b\langle0|\otimes (U_{B_2}U_{B_1})+|1\rangle_b\langle1|\otimes (U_{B_1}U_{B_2}).
\end{split}
\end{eqnarray}

Alice and Bob measure the two control qubits $a$ and $b$ in the basis $\{|\pm\rangle\}$, respectively. If the measurement results are $|+\rangle_a|+\rangle_b$ or $|-\rangle_a|-\rangle_b$, the state $|\Phi_1'\rangle$ will collapse to a normalized state
\begin{eqnarray}
\begin{split}             \label{eq9}
|\Phi_2\rangle=&\;\frac{1}{\sqrt{2}}[(U_{A_2}U_{A_1})\otimes(U_{B_2}U_{B_1})\\&
                                     +(U_{A_1}U_{A_2})\otimes(U_{B_1}U_{B_2})]|\psi\rangle_A|\psi\rangle_B,
\end{split}
\end{eqnarray}
with a success probability of $\frac{1}{2}$. If the measurement results are $|+\rangle_a|-\rangle_b$ or $|-\rangle_a|+\rangle_b$, the state $|\Phi_1'\rangle$ will collapse to a normalized state
\begin{eqnarray}
\begin{split}             \label{eq10}
|\Phi_2'\rangle=&\;\frac{1}{\sqrt{2}}[(U_{A_2}U_{A_1})\otimes(U_{B_2}U_{B_1})\\&
                                     -(U_{A_1}U_{A_2})\otimes(U_{B_1}U_{B_2})]|\psi\rangle_A|\psi\rangle_B,
\end{split}
\end{eqnarray}
with a success probability of $\frac{1}{2}$.

More generally,  in the above approach, operations
$\frac{1}{\sqrt{2}}[(U_{A_2}U_{A_1})\otimes(U_{B_2}U_{B_1})\pm(U_{A_1}U_{A_2})\otimes(U_{B_1}U_{B_2})]$ are applied to transform the initial product state $|\psi\rangle_A\otimes|\psi\rangle_B$ into a MES, PES, or EE. To achieve this aim, we introduce the following single-qubit gates $U_{A_1}, U_{A_2}, U_{B_1}$, and $U_{B_2}$:
\begin{eqnarray}
\begin{split}             \label{eq11}
&U_{A_2}=X = \left(\begin{array}{cc}
             0  &   1  \\
             1  &   0  \\
\end{array}\right),\\
&U_{B_2}=X = \left(\begin{array}{cc}
             0  &   1  \\
             1  &   0  \\
\end{array}\right),\\
&U_{A_1}=R_y(2\theta)=\left(\begin{array}{cc}
           \cos\theta  &  -\sin\theta  \\
           \sin\theta  &  \cos\theta  \\
\end{array}\right), \\
&U_{B_1}=R_y(2\theta)=\left(\begin{array}{cc}
           \cos\theta  &  -\sin\theta  \\
           \sin\theta  &  \cos\theta  \\
\end{array}\right).
\end{split}
\end{eqnarray}
Here $\theta\in[0, 2\pi]$.
Substituting Eq. \eqref{eq11} into Eq. \eqref{eq9} and Eq. \eqref{eq10}, the states $|\Phi_2\rangle$ and $|\Phi_2'\rangle$ can be rewritten as
\begin{eqnarray}
\begin{split}             \label{eq12}
|\Phi_2\rangle=&\;\sqrt{2}[(\sqrt{(1-\alpha_A)(1-\alpha_B)}\cos^2\theta \\& + \sqrt{\alpha_A \alpha_B}\sin^2\theta)|0\rangle_A|0\rangle_B
\\&+(\sqrt{(1-\alpha_A)\alpha_B}\cos^2\theta \\& - \sqrt{\alpha_A(1-\alpha_B)}   \sin^2\theta) |0\rangle_A|1\rangle_B  %
\\&+(\sqrt{\alpha_A(1-\alpha_B)}\cos^2\theta \\& - \sqrt{(1-\alpha_A)\alpha_B}   \sin^2\theta) |1\rangle_A|0\rangle_B
\\&+(\sqrt{\alpha_A\alpha_B}    \cos^2\theta \\& + \sqrt{(1-\alpha_A)(1-\alpha_B)}\sin^2\theta)|1\rangle_A|1\rangle_B].
\end{split}
\end{eqnarray}
\begin{eqnarray}
\begin{split}             \label{eq13}
|\Phi_2'\rangle=&\;\frac{1}{\sqrt{2}}\sin2\theta[(\sqrt{\alpha_A(1-\alpha_B)}+\sqrt{(1-\alpha_A)\alpha_B})\\&\times(|0\rangle_A|0\rangle_B-|1\rangle_A|1\rangle_B)
                          +(\sqrt{\alpha_A\alpha_B}\\&-\sqrt{(1-\alpha_A)(1-\alpha_B)})(|0\rangle_A|1\rangle_B + |1\rangle_A|0\rangle_B)].
\end{split}
\end{eqnarray}
%
%

As shown in Table \ref{table1},  based on Eqs. \eqref{eq12}-\eqref{eq13} and the specific choice of parameters, the measurement results, output states, and the types of output states can be classified as follows:

(i) If the measurement results are $|+\rangle_a|+\rangle_b$ or $|-\rangle_a|-\rangle_b$, the output state will be $|\Phi_2\rangle$.
(a) When $\theta=\frac{\pi}{4}$, $\frac{3\pi}{4}$, $\frac{5\pi}{4}$, $\frac{7\pi}{4}$ with $\alpha_A=\alpha_B=\frac{1}{2}$ or 0, or 1, the state $|\Phi_2\rangle$ will become a MES, i.e., $|\Phi_2\rangle=|\Phi^+\rangle=\frac{1}{\sqrt{2}}(|0\rangle_A|0\rangle_B+|1\rangle_A|1\rangle_B)$.
(b) When $\theta=\frac{\pi}{4}$, $\frac{3\pi}{4}$, $\frac{5\pi}{4}$, $\frac{7\pi}{4}$ with $\alpha_A=0$ and $\alpha_B=1$, the state $|\Phi_2\rangle$ will become a MES, i.e.,  $|\Phi_2\rangle=|\Psi^-\rangle=\frac{1}{\sqrt{2}}(|0\rangle_A|1\rangle_B-|1\rangle_A|0\rangle_B)$.
(c) When $\theta=\frac{\pi}{4}$, $\frac{3\pi}{4}$, $\frac{5\pi}{4}$, $\frac{7\pi}{4}$ with $\alpha_A=1$ and $\alpha_B=0$, the state $|\Phi_2\rangle$ will become a MES, i.e.,  $|\Phi_2\rangle=-|\Psi^-\rangle$.

(ii) If the measurement results are $|+\rangle_a|-\rangle_b$ or $|-\rangle_a|+\rangle_b$, the output state will be $|\Phi_2'\rangle$.
(a) When $\theta=\frac{\pi}{4}$, $\frac{3\pi}{4}$, $\frac{5\pi}{4}$, $\frac{7\pi}{4}$ with $\alpha_A=\alpha_B=\frac{1}{2}$ (or $\alpha_A=0$, $\alpha_B=1$, or $\alpha_A=1$, $\alpha_B=0$), the state $|\Phi_2'\rangle$ will become a MES, i.e., $|\Phi_2'\rangle=\pm|\Phi^-\rangle=\pm\frac{1}{\sqrt{2}}(|0\rangle_A|0\rangle_B-|1\rangle_A|1\rangle_B)$.
(b) When $\theta=\frac{\pi}{4}$, $\frac{3\pi}{4}$, $\frac{5\pi}{4}$, $\frac{7\pi}{4}$ with $\alpha_A=\alpha_B=0$  (or $\alpha_A=\alpha_B=1$), the state $|\Phi_2'\rangle$ will become a MES, i.e., $|\Phi_2'\rangle=\pm|\Psi^+\rangle=\pm\frac{1}{\sqrt{2}}(|0\rangle_A|1\rangle_B+|1\rangle_A|0\rangle_B)$.

(iii) If $\theta=0$, $\pi$, $2\pi$ or $\theta=\frac{\pi}{2}$, $\frac{3\pi}{2}$, no matter what the values of $\alpha_A$ and $\alpha_B$ are taken, no entanglement state can be obtained. Specifically,
(a) when $\theta=0$, $\pi$, $2\pi$, the state
$|\Phi_2\rangle$ will become a SS, i.e., $|\Phi_2\rangle=|\Phi^{xx}\rangle=\sigma_x^{A}\sigma_x^{B}(|\psi\rangle_A\otimes|\psi\rangle_B)$,
while $|\Phi_2'\rangle$ will vanish.
(b) When $\theta=\frac{\pi}{2}$, $\frac{3\pi}{2}$, the state
$|\Phi_2\rangle$ will become a SS, i.e., $|\Phi_2\rangle=|\Phi^{zz}\rangle=\sigma_z^{A}\sigma_z^{B}(|\psi\rangle_A\otimes|\psi\rangle_B)$, while
$|\Phi_2'\rangle$ will vanish.
Here $\sigma_x^{A}$ ($\sigma_x^{B}$) and $\sigma_z^{A}$ ($\sigma_z^{B}$) are the Pauli $X$ and  Pauli $Z$ gates acting on qubit $A$ ($B$).

(iv) If the coefficients $\alpha_A=\alpha_B=0$ (or $\alpha_A=\alpha_B=1$) with $\theta \neq\frac{k\pi}{4} (k=0,1,\ldots,8)$, the output state $|\Phi_2\rangle=\sqrt{2}(\cos^2\theta|0\rangle_A|0\rangle_B+\sin^2\theta|1\rangle_A|1\rangle_B)$ (or $|\Phi_2\rangle=\sqrt{2}(\sin^2\theta|0\rangle_A|0\rangle_B+\cos^2\theta|1\rangle_A|1\rangle_B)$) is a less-entangled Bell state, i.e., a PES. While $|\Phi_2'\rangle$ can be a MES for $\theta \neq 0, \frac{\pi}{2}, \pi, \frac{3\pi}{2}, 2\pi$ and $\alpha_A=\alpha_B=0$ (or $\alpha_A=\alpha_B=1$).

(v) For all other cases, the output states $|\Phi_2\rangle$ and $|\Phi_2'\rangle$ both are PES.
For example, when $\alpha_A=\frac{1}{2}$, $\alpha_B=\frac{1}{3}$, and $\theta=\frac{\pi}{4}$,
$|\Phi_2\rangle=\frac{1+\sqrt{2}}{2\sqrt{3}}(|0\rangle_A|0\rangle_B+|1\rangle_A|1\rangle_B)+\frac{1-\sqrt{2}}{2\sqrt{3}}(|0\rangle_A|1\rangle_B-|1\rangle_A|0\rangle_B)$  and $|\Phi_2'\rangle=\frac{1+\sqrt{2}}{2\sqrt{3}}(|0\rangle_A|0\rangle_B-|1\rangle_A|1\rangle_B)+\frac{1-\sqrt{2}}{2\sqrt{3}}(|0\rangle_A|1\rangle_B+|1\rangle_A|0\rangle_B)$ both are PES. 

Therefore, the protocol illustrated in Fig. \ref{Switch} deterministically generates a two-qubit MES on two long-distance quibts $A$ and $B$. The protocol success is heralded by the measurement outcomes of qubits $a$ and $b$. If we wish to obtain the same output states, two-bit classical communication between Alice and Bob, along with some feed-forward operations on qubits $A$ and $B$, will be required.

\begin{table}[tb]
\centering
\caption{The relations between the angle $\theta$ described in Eq. \eqref{eq11}, the measurement results of qubits $a$ and $b$,  and the output states $|\Phi_2\rangle$ and $|\Phi_2'\rangle$.
$\theta\in[0, 2\pi]$ and $n=1, 2, 3, 4$.
$|\Phi^\pm\rangle=\frac{1}{\sqrt{2}}(|0\rangle_A|0\rangle_B\pm|1\rangle_A|1\rangle_B)$, $|\Psi^\pm\rangle=\frac{1}{\sqrt{2}}(|0\rangle_A|1\rangle_B\pm|1\rangle_A|0\rangle_B)$,
$|\Phi^{xx}\rangle=\sigma_x^{A}\sigma_x^{B}(|\psi\rangle_A\otimes|\psi\rangle_B)$, and
$|\Phi^{zz}\rangle=\sigma_z^{A}\sigma_z^{B}(|\psi\rangle_A\otimes|\psi\rangle_B)$,
where $\sigma_x^{A}$ ($\sigma_x^{B}$) and $\sigma_z^{A}$ ($\sigma_z^{B}$) are the Pauli $X$ and  Pauli $Z$ gates acting on qubit $A$ ($B$).}
\begin{tabular}{ccccccc}
\hline\hline
$\theta$               &   $\alpha_A$  &  $\alpha_B$  &  Measured result &  Output state &    Type \\
\hline
$\frac{(2n-1)\pi}{4}$  &   $\frac{1}{2}$ &  $\frac{1}{2}$ & \makecell*[c]{$|+\rangle_a|+\rangle_b, \, |-\rangle_a|-\rangle_b$ \\   $|+\rangle_a|-\rangle_b, \, |-\rangle_a|+\rangle_b$}  & \makecell*[c]{$|\Phi_2\rangle=|\Phi^+\rangle$  \\ $|\Phi_2'\rangle=\pm|\Phi^-\rangle$} & MES \\
$\frac{(2n-1)\pi}{4}$  &  0            &  0            & \makecell*[c]{ $|+\rangle_a|+\rangle_b, \, |-\rangle_a|-\rangle_b$ \\ $|+\rangle_a|-\rangle_b, \, |-\rangle_a|+\rangle_b$} &  \makecell*[c]{$|\Phi_2\rangle=|\Phi^+\rangle$ \\ $|\Phi_2'\rangle=\pm|\Psi^+\rangle$} & MES \\
$\frac{(2n-1)\pi}{4}$  &  1            &  1            & \makecell*[c]{ $|+\rangle_a|+\rangle_b, \, |-\rangle_a|-\rangle_b$ \\ $|+\rangle_a|-\rangle_b, \,  |-\rangle_a|+\rangle_b$} &  \makecell*[c]{ $|\Phi_2\rangle=|\Phi^+\rangle$ \\ $|\Phi_2'\rangle=\pm|\Psi^+\rangle$} & MES \\
$\frac{(2n-1)\pi}{4}$  &  0            &  1            &\makecell*[c]{ $|+\rangle_a|+\rangle_b, \, |-\rangle_a|-\rangle_b$ \\ $|+\rangle_a|-\rangle_b, \, |-\rangle_a|+\rangle_b$} &  \makecell*[c]{$|\Phi_2\rangle=|\Psi^-\rangle$ \\ $|\Phi_2'\rangle=\pm|\Phi^-\rangle$} & MES \\
$\frac{(2n-1)\pi}{4}$  &  1            & 0            & \makecell*[c]{$|+\rangle_a|+\rangle_b, \, |-\rangle_a|-\rangle_b$ \\ $|+\rangle_a|-\rangle_b, \, |-\rangle_a|+\rangle_b$} &  \makecell*[c]{ $|\Phi_2\rangle=-|\Psi^-\rangle$ \\ $|\Phi_2'\rangle=\pm|\Phi^-\rangle$} & MES \\
$0, \pi, 2\pi$  & Any             & Any            & \makecell*[c]{$|+\rangle_a|+\rangle_b, \, |-\rangle_a|-\rangle_b$ \\ $|+\rangle_a|-\rangle_b, \, |-\rangle_a|+\rangle_b$} &  \makecell*[c]{ $|\Phi_2\rangle=|\Phi^{xx}\rangle$ \\ $|\Phi_2'\rangle=\textbf{0}$} &  \makecell*[c]{ SS \\ \textbf{0} state} \\
$\frac{\pi}{2}, \frac{3\pi}{2}$  & Any            & Any           & \makecell*[c]{$|+\rangle_a|+\rangle_b, \, |-\rangle_a|-\rangle_b$ \\ $|+\rangle_a|-\rangle_b, \, |-\rangle_a|+\rangle_b$} &  \makecell*[c]{ $|\Phi_2\rangle=|\Phi^{zz}\rangle$ \\ $|\Phi_2'\rangle=\textbf{0}$} &  \makecell*[c]{ SS \\ \textbf{0} state} \\
\hline\hline
\end{tabular}
\label{table1}
\end{table}

\subsection{Entangled state generation among three long-distance qubits}  \label{Sec2.2}


\begin{figure} 
\begin{center}
\includegraphics[width=8 cm,angle=0]{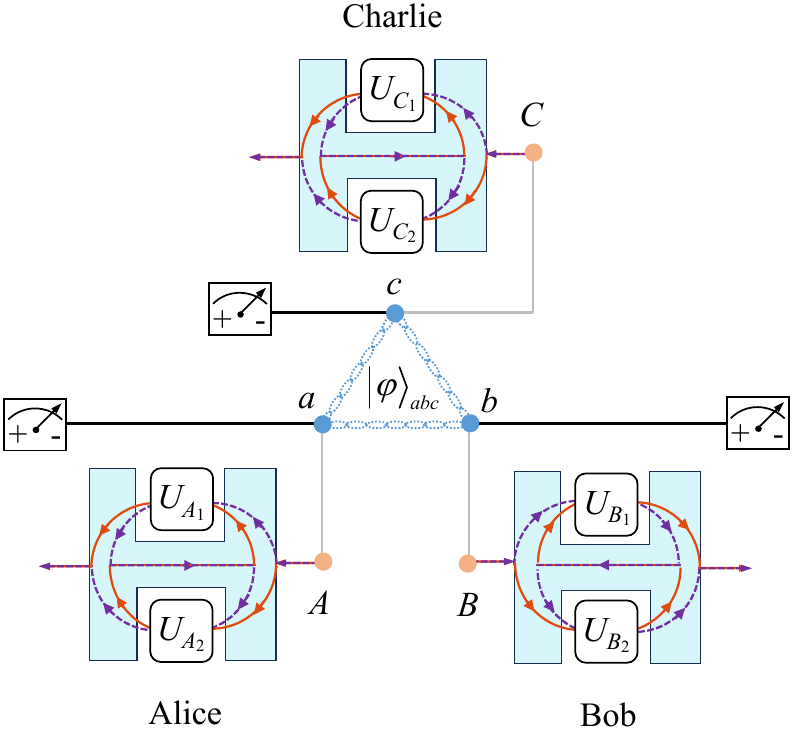}
\caption{Schematic diagram for generating a  heralded general three-qubit quantum state on three remote particles $A$, $B$, and $C$ via three quantum switches.
First, a maximally entangled state $|\varphi\rangle_{abc}$ is shared with Alice, Bob, and Charlie.
Qubits $a$, $b$, and $c$ act as the control qubit that manipulate the operation order on target qubits $A$, $B$, and $C$ via their individual quantum switches, respectively.
When the control qubit $a$ is in the state $|0\rangle_a$, the operation order of single-qubit gate is that $U_{A_1}$ is before $U_{A_2}$, i.e., $U_{A_2}U_{A_1}$, denoted by the purple circuit.
When the control qubit $a$ is in the state $|1\rangle_a$, the operation order of single-qubit gate is that $U_{A_2}$ is before $U_{A_1}$, i.e., $U_{A_1}U_{A_2}$, denoted by the red circuit.
Bob and Charlie come to the same conclusion.
Finally, qubits $a$, $b$, and $c$ are measured in the basis $\{|\pm\rangle\}$, the state $ABC$ will become a MES, PES, or SS, by appropriately introducing the single-qubit gates $U_{A_1}$, $U_{A_2}$, $U_{B_1}$, $U_{B_2}$, $U_{C_1}$, and $U_{C_2}$.}
\label{Switch2}
\end{center}
\end{figure}

Suppose that three parties Alice, Bob, and Charlie possess particles $A$, $B$, and $C$, respectively. These particles are initially prepared as
\begin{eqnarray}            \label{eq14}
|\psi\rangle_A =\sqrt{\alpha_A} |0\rangle_A+\sqrt{1-\alpha_A} |1\rangle_A,
\end{eqnarray}
\begin{eqnarray}            \label{eq15}
|\psi\rangle_B =\sqrt{\alpha_B} |0\rangle_B+\sqrt{1-\alpha_B} |1\rangle_B,
\end{eqnarray}
\begin{eqnarray}            \label{eq16}
|\psi\rangle_C =\sqrt{\alpha_C} |0\rangle_C+\sqrt{1-\alpha_C} |1\rangle_C.
\end{eqnarray}
The objective of Alice, Bob, and Charlie is to transform the given  separable state $|\psi\rangle_A\otimes|\psi\rangle_B\otimes |\psi\rangle_C$ into a general three-qubit state.

As shown in Fig. \ref{Switch2}, we propose a heralded scheme for the deterministic implementation of a general long-distance three-qubit state. The detailed procedure of our protocol is outlined step by step below.

Step 1: A three-qubit maximally entangled state
\begin{eqnarray}            \label{eq17}
|\varphi\rangle_{abc}=\frac{1}{\sqrt{2}}(|0\rangle_{a}|0\rangle_{b}|0\rangle_{c}+|1\rangle_{a}|1\rangle_{b}|1\rangle_{c}),
\end{eqnarray}
is shared with Alice, Bob, and Charlie. Here particles $a$, $b$, and $c$ are held by Alice, Bob, and Charlie, respectively.
The state of the whole system is given by
\begin{eqnarray}
\begin{split}                \label{eq18}
|\Psi\rangle_{0}= &\;|\varphi\rangle_{abc}\otimes|\psi\rangle_A\otimes|\psi\rangle_B\otimes|\psi\rangle_C\\
                =&\;\frac{1}{\sqrt{2}}(|0\rangle_a|0\rangle_b|0\rangle_c + |1\rangle_a|1\rangle_b|1\rangle_c)\\&\otimes|\psi\rangle_A\otimes|\psi\rangle_B\otimes|\psi\rangle_C.
\end{split}
\end{eqnarray}
\begin{table*}[tb]
\centering
\caption{The correspondence between the values of the rotation angle $\theta$, the coefficients $\alpha_A$, $\alpha_B$, and $\alpha_C$, the measured results on qubits $a$, $b$, and $c$, the final output states on qubits $A$, $B$, and $C$, and the state type of the output states.
$|\Phi_3^\pm\rangle=\frac{1}{\sqrt{2}}(|0\rangle_A|0\rangle_B|0\rangle_C\pm|1\rangle_A|1\rangle_B|1\rangle_C)$, $|\Phi^{xxx}\rangle=\sigma_x^{A}\sigma_x^{B}\sigma_x^{C}(|\psi\rangle_A\otimes|\psi\rangle_B\otimes|\psi\rangle_C)$, and $|\Phi^{zzz}\rangle=\sigma_z^{A}\sigma_z^{B}\sigma_z^{C}(|\psi\rangle_A\otimes|\psi\rangle_B\otimes|\psi\rangle_C)$.}

\begin{tabular}{cccccccc}
\hline\hline
$\theta$              &\quad $\alpha_A$       &\quad $\alpha_B$       & \quad $\alpha_C$      & \quad Measured result   &\quad Output state     & \quad State type \\
\hline
$\frac{(2n-1)\pi}{4}$ &\quad $\frac{1}{2}$    &\quad  $\frac{1}{2}$    &\quad   $\frac{1}{2}$   & \qquad \makecell*[c]{ $|+\rangle_a|+\rangle_b|+\rangle_c, \, |+\rangle_a|-\rangle_b|-\rangle_c, \, |-\rangle_a|+\rangle_b|-\rangle_c, \, |-\rangle_a|-\rangle_b|+\rangle_c $ \\ $|+\rangle_a|+\rangle_b|-\rangle_c, \, |+\rangle_a|-\rangle_b|+\rangle_c, \, |-\rangle_a|+\rangle_b|+\rangle_c, \, |-\rangle_a|-\rangle_b|-\rangle_c $}                & \quad  \makecell*[c]{$|\Psi_2\rangle=|\Phi^+_3\rangle$\\ $|\Psi_2'\rangle=|\Phi_3^-\rangle$} & \quad MES \\
$0, \pi, 2\pi $          &\quad  Any              &\quad Any               &\quad  Any             & \qquad \makecell*[c]{ $|+\rangle_a|+\rangle_b|+\rangle_c, \, |+\rangle_a|-\rangle_b|-\rangle_c, \, |-\rangle_a|+\rangle_b|-\rangle_c, \, |-\rangle_a|-\rangle_b|+\rangle_c $ \\ $|+\rangle_a|+\rangle_b|-\rangle_c, \, |+\rangle_a|-\rangle_b|+\rangle_c, \, |-\rangle_a|+\rangle_b|+\rangle_c, \, |-\rangle_a|-\rangle_b|-\rangle_c $}                  &\quad  \makecell*[c]{$|\Psi_2\rangle=|\Phi^{xxx}\rangle$\\ $|\Psi_2'\rangle=\textbf{0}$} &\quad  \makecell*[c]{SS\\ \textbf{0} state}  \\
$\frac{\pi}{2}, \frac{3\pi}{2} $          &\quad Any              &\quad Any               &\quad  Any             &\qquad  \makecell*[c]{ $|+\rangle_a|+\rangle_b|+\rangle_c, \, |+\rangle_a|-\rangle_b|-\rangle_c, \, |-\rangle_a|+\rangle_b|-\rangle_c, \, |-\rangle_a|-\rangle_b|+\rangle_c $ \\ $|+\rangle_a|+\rangle_b|-\rangle_c, \, |+\rangle_a|-\rangle_b|+\rangle_c, \, |-\rangle_a|+\rangle_b|+\rangle_c, \, |-\rangle_a|-\rangle_b|-\rangle_c $}                  &\quad  \makecell*[c]{$|\Psi_2\rangle=\textbf{0}$\\ $|\Psi_2'\rangle=|\Phi^{zzz}\rangle$} &\quad  \makecell*[c]{\textbf{0} state \\ SS}  \\
\hline\hline
\end{tabular}
\label{table2}
\end{table*}

Step 2: A similar arrangement as that made in Sec. \ref{Sec2.1}.  Qubits $a$, $b$ and $c$ act as the control qubits to govern the order of single-qubit gates on qubits $A$, $B$, and $C$ by resorting to their individual quantum switches.
Here each quantum switch non-trivially performs the gate operations in the order $U_{i_2}U_{i_1}$ ($i=A$, $B$, and $C$) when the control qubit is in the state $|0\rangle$,
                             while the gate operations are performed in the order $U_{i_1}U_{i_2}$ ($i=A$, $B$, and $C$) when the control qubit is in the state $|1\rangle$.
Thus, three quantum switches change the initial state $|\Psi\rangle_{0}$ into
\begin{eqnarray}
\begin{split}                \label{eq19}
|\Psi_1\rangle =&\;\frac{1}{\sqrt{2}}[|0\rangle_a (U_{A_2}U_{A_1}|\psi\rangle_A)\otimes |0\rangle_b (U_{B_2}U_{B_1}|\psi\rangle_B) \\&
\otimes |0\rangle_c (U_{C_2}U_{C_1}|\psi\rangle_C)
                + |1\rangle_a (U_{A_1}U_{A_2}|\psi\rangle_A)\\&
                \otimes |1\rangle_b (U_{B_1}U_{B_2}|\psi\rangle_B)\otimes |1\rangle_c (U_{C_1}U_{C_2}|\psi\rangle_C)].
\end{split}
\end{eqnarray}

Step 3: Alice, Bob, and Charlie measure the control qubits $a$, $b$, and $c$ in the basis $\{|\pm\rangle\}$, respectively.
If the measured results are $|+\rangle_a|+\rangle_b|+\rangle_c$, $|+\rangle_a|-\rangle_b|-\rangle_c$, $|-\rangle_a|+\rangle_b|-\rangle_c$, or $|-\rangle_a|-\rangle_b|+\rangle_c$, the state $|\Psi_1\rangle$ will collapse to a normalized state
\begin{eqnarray}
\begin{split}                \label{eq20}
|\Psi_2\rangle =&\;\frac{1}{\sqrt{2}}[(U_{A_2}U_{A_1})\otimes(U_{B_2}U_{B_1})\otimes(U_{C_2}U_{C_1})\\&
                                     +(U_{A_1}U_{A_2})\otimes(U_{B_1}U_{B_2})\otimes(U_{C_1}U_{C_2})](|\psi\rangle_A\\&\otimes|\psi\rangle_B\otimes|\psi\rangle_C),
\end{split}
\end{eqnarray}
with a success probability of $\frac{1}{2}$. If the measured results are $|+\rangle_a|+\rangle_b|-\rangle_c$,  $|+\rangle_a|-\rangle_b|+\rangle_c$,  $|-\rangle_a|+\rangle_b|+\rangle_c$, or $|-\rangle_a|-\rangle_b|-\rangle_c$, then the state $|\Psi_1\rangle$ will collapse into
\begin{eqnarray}
\begin{split}                \label{eq21}
|\Psi_2'\rangle =&\;\frac{1}{\sqrt{2}}[(U_{A_2}U_{A_1})\otimes(U_{B_2}U_{B_1})\otimes(U_{C_2}U_{C_1})\\&
                                     -(U_{A_1}U_{A_2})\otimes(U_{B_1}U_{B_2})\otimes(U_{C_1}U_{C_2})]\\&\otimes(|\psi\rangle_A\otimes|\psi\rangle_B\otimes|\psi\rangle_C),
\end{split}
\end{eqnarray}
with a probability of $\frac{1}{2}$. To obtain a general quantum state of remote qubits $A$, $B$, and $C$, we introduce
%
\begin{eqnarray}
\begin{split}             \label{eq22}
&U_{A_2}=U_{B_2}=U_{C_2}=X, \\&
U_{A_1}=U_{B_1}=U_{C_1}=R_y(2\theta).
\end{split}
\end{eqnarray}
%
Then, $|\Psi_2\rangle$ and $|\Psi_2'\rangle$ can be reexpressed as:
\begin{eqnarray}
\begin{split}                \label{eq23}
|\Psi_2\rangle =&\;\frac{1}{\sqrt{2}}[(\alpha_2\sin\theta\sin2\theta
                                      +\alpha_3\sin\theta\sin2\theta\\&
                                      +\alpha_5\sin\theta\sin2\theta
                                      +2\alpha_8\cos^3\theta) |0\rangle_A|0\rangle_B|0\rangle_C\\&
                                      +(\alpha_1\sin\theta\sin2\theta
                                      -\alpha_4\sin\theta\sin2\theta\\&
                                      -\alpha_6\sin\theta\sin2\theta
                                      +2\alpha_7\cos^3\theta) |0\rangle_A|0\rangle_B|1\rangle_C\\&
                                      +(\alpha_1\sin\theta\sin2\theta
                                      -\alpha_4\sin\theta\sin2\theta\\&
                                      +2\alpha_6 \cos^3\theta
                                      -\alpha_7\sin\theta\sin2\theta) |0\rangle_A|1\rangle_B|0\rangle_C\\&
                                      -(\alpha_2\sin\theta\sin2\theta
                                      +\alpha_3\sin\theta\sin2\theta\\&
                                      -2\alpha_5 \cos^3\theta
                                      -\alpha_8\sin\theta\sin2\theta) |0\rangle_A|1\rangle_B|1\rangle_C\\&
                                      +(\alpha_1\sin\theta\sin2\theta
                                      +2\alpha_4\cos^3\theta\\&
                                      -\alpha_6\sin\theta\sin2\theta
                                      -\alpha_7\sin\theta\sin2\theta) |1\rangle_A|0\rangle_B|0\rangle_C\\&
                                      -(\alpha_2\sin\theta\sin2\theta
                                      -2\alpha_3\cos^3\theta\\&
                                      +\alpha_5\sin\theta\sin2\theta
                                      -\alpha_8\sin\theta\sin2\theta) |1\rangle_A|0\rangle_B|1\rangle_C\\&
                                      +(2\alpha_2\cos^3\theta
                                      -\alpha_3\sin\theta\sin2\theta\\&
                                      -\alpha_5\sin\theta\sin2\theta
                                      +\alpha_8\sin\theta\sin2\theta) |1\rangle_A|1\rangle_B|0\rangle_C\\&
                                      +(2\alpha_1\cos^3\theta
                                      +\alpha_4\sin\theta\sin2\theta\\&
                                      +\alpha_6\sin\theta\sin2\theta
                                      +\alpha_7\sin\theta\sin2\theta) |1\rangle_A|1\rangle_B|1\rangle_C].
\end{split}
\end{eqnarray}
\begin{eqnarray}
\begin{split}                \label{eq24}
|\Psi_2'\rangle =&\;\frac{1}{\sqrt{2}}[(2\alpha_1\sin^3\theta
                                         +\alpha_4\cos\theta\sin2\theta\\&
                                         +\alpha_6\cos\theta\sin2\theta
                                         +\alpha_7\cos\theta\sin2\theta) |0\rangle_A|0\rangle_B|0\rangle_C\\&
                                         -(2\alpha_2\sin^3\theta
                                         -\alpha_3\cos\theta\sin2\theta\\&
                                         -\alpha_5\cos\theta\sin2\theta
                                         +\alpha_8\cos\theta\sin2\theta) |0\rangle_A|0\rangle_B|1\rangle_C\\&
                                         +(\alpha_2\cos\theta\sin2\theta
                                         -2\alpha_3\sin^3\theta\\&
                                         +\alpha_5 \cos\theta\sin2\theta
                                         -\alpha_8\cos\theta\sin2\theta) |0\rangle_A|1\rangle_B|0\rangle_C\\&
                                         +(\alpha_1\cos\theta\sin2\theta
                                         -\alpha_6\cos\theta\sin2\theta\\&
                                         +2\alpha_4\sin^3\theta
                                         -\alpha_7\cos\theta\sin2\theta) |0\rangle_A|1\rangle_B|1\rangle_C\\&
                                         +(\alpha_2\cos\theta\sin2\theta
                                         +\alpha_3\cos\theta\sin2\theta\\&
                                         -2\alpha_5\sin^3\theta
                                         -\alpha_8\cos\theta\sin2\theta) |1\rangle_A|0\rangle_B|0\rangle_C\\&
                                         +(\alpha_1\cos\theta\sin2\theta
                                         -\alpha_4\cos\theta\sin2\theta\\&
                                         +2\alpha_6\sin^3\theta
                                         -\alpha_7\cos\theta\sin2\theta) |1\rangle_A|0\rangle_B|1\rangle_C\\&
                                         +(\alpha_1\cos\theta\sin2\theta
                                         -\alpha_4\cos\theta\sin2\theta\\&
                                         -\alpha_6\cos\theta\sin2\theta
                                         +2\alpha_7\sin^3\theta)  |1\rangle_A|1\rangle_B|0\rangle_C\\&
                                         -(\alpha_2\cos\theta\sin2\theta
                                         +\alpha_3\cos\theta\sin2\theta\\&
                                         +\alpha_5\cos\theta\sin2\theta
                                         +2\alpha_8\sin^3\theta)  |1\rangle_A|1\rangle_B|1\rangle_C].
\end{split}
\end{eqnarray}
For convenience, we here denote the coefficients as
\begin{eqnarray}
\begin{split}                \label{eq}
\alpha_1 =&\;\sqrt{\alpha_A\alpha_B\alpha_C}, \\
\alpha_2 =&\;\sqrt{\alpha_A\alpha_B(1-\alpha_C)},\\
\alpha_3 =&\;\sqrt{\alpha_A(1-\alpha_B)\alpha_C},\\
\alpha_4 =&\;\sqrt{\alpha_A(1-\alpha_B)(1-\alpha_C)},\\
\alpha_5 =&\;\sqrt{(1-\alpha_A)\alpha_B\alpha_C},  \\
\alpha_6 =&\;\sqrt{(1-\alpha_A)\alpha_B(1-\alpha_C)}, \\
\alpha_7 =&\;\sqrt{(1-\alpha_A)(1-\alpha_B)\alpha_C}, \\
\alpha_8 =&\;\sqrt{(1-\alpha_A)(1-\alpha_B)(1-\alpha_C)}.
\end{split}
\end{eqnarray}

Table \ref{table2} presents the relations between the values of parameters, measured results, the final output states, and state type of output states on qubits $A$, $B$, and $C$.  As shown in Tab. \ref{table2}, the output states can be classified into four groups: MES, SS, vanished state, and PES.

(i) In the MES group, when the angle $\theta=\frac{\pi}{4}$, $\frac{3\pi}{4}$, $\frac{5\pi}{4}$, $\frac{7\pi}{4}$ with $\alpha_A=\alpha_B=\alpha_C=\frac{1}{2}$ are taken, the MESs $|\Psi_2\rangle=\frac{1}{\sqrt{2}}(|0\rangle_A|0\rangle_B|0\rangle_C+|1\rangle_A|1\rangle_B|1\rangle_C)$ and $|\Psi_2'\rangle=\frac{1}{\sqrt{2}}(|0\rangle_A|0\rangle_B|0\rangle_C-|1\rangle_A|1\rangle_B|1\rangle_C)$ are obtained.

(ii) In the SS group,
when $\theta=0$, $\pi$, or $2\pi$, the state $|\Psi_2\rangle$ becomes a coefficient-independent SS $|\Psi_2\rangle=|\Phi^{xxx}\rangle=\sigma_x^{A}\sigma_x^{B}\sigma_x^{C}(|\psi\rangle_A\otimes|\psi\rangle_B\otimes|\psi\rangle_C)$. Similarly, when $\theta=\frac{\pi}{2}$ or $\frac{3\pi}{2}$, the state $|\Psi_2'\rangle$ becomes a coefficient-independent SS $|\Psi_2'\rangle=|\Phi^{zzz}\rangle=\sigma_z^{A}\sigma_z^{B}\sigma_z^{C}(|\psi\rangle_A\otimes|\psi\rangle_B\otimes|\psi\rangle_C)$.

(iii) In the vanished state group,
when $\theta=0$, $\pi$, or $2\pi$, the state $|\Psi_2'\rangle=\textbf{0}$ will be vanished for all possible coefficients $\alpha_A, \alpha_B$, and $\alpha_C$. Similarly,
when $\theta=\frac{\pi}{2}$ or $\frac{3\pi}{2}$,  the state $|\Psi_2\rangle=\textbf{0}$ will be vanished for all possible coefficients $\alpha_A, \alpha_B$, and $\alpha_C$.

(iv) In the PES group, for all other cases, the output states $|\Psi_2\rangle$ and $|\Psi_2'\rangle$ will both be PES. For example, when $\alpha_A=\alpha_B=\alpha_C=0$ with $\theta\neq0,\frac{\pi}{2},\pi,\frac{3\pi}{2},2\pi$, $|\Psi_2\rangle=\frac{1}{\sqrt{2}}[2\cos^3\theta|0\rangle_A|0\rangle_B|0\rangle_C+\sin\theta\sin2\theta(|1\rangle_A|1\rangle_B|0\rangle_C-|0\rangle_A|1\rangle_B|1\rangle_C-|1\rangle_A|0\rangle_B|1\rangle_C)]$ and $|\Psi_2'\rangle=\frac{1}{\sqrt{2}}[2\sin^3\theta|1\rangle_A|1\rangle_B|1\rangle_C+\cos\theta\sin2\theta(|0\rangle_A|0\rangle_B|1\rangle_C-|0\rangle_A|1\rangle_B|0\rangle_C-|1\rangle_A|0\rangle_B|0\rangle_C)]$ both are PES. Specifically, when $\theta=\frac{\pi}{4}$, $|\Psi_2\rangle$ and $|\Psi_2'\rangle$ will become GHZ-like states \cite{dong2011controlled}.

\subsection{Entangled state generation among $N$ separable long-distance qubits}  \label{Sec2.3}

\begin{figure} 
\begin{center}
\includegraphics[width=7.8 cm,angle=0]{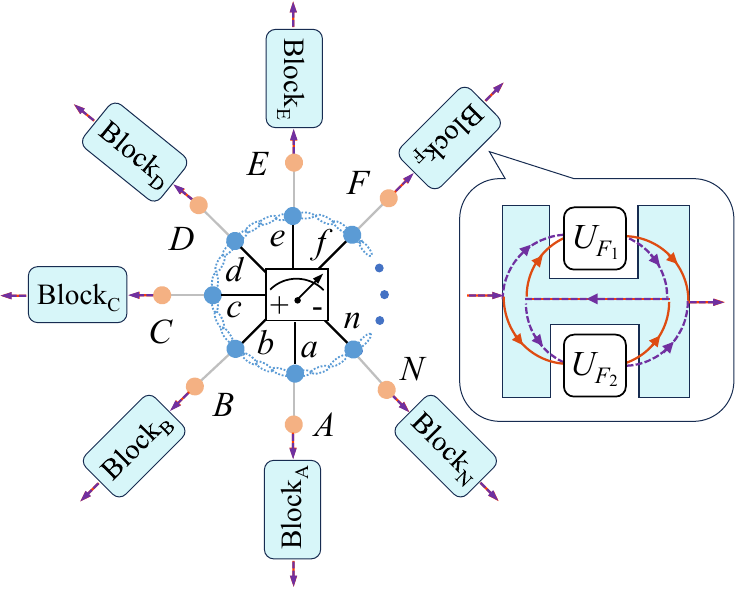}
\caption{Schematic diagram of a scheme designed to generate a heralded $N$-qubit entangled state on $N$ remote separable qubits $A$ to $N$ via $N$ quantum switches. Here each ``Block" represents a quantum switch. First, a maximally entangled state $|\varphi\rangle_{ab\ldots n}=\frac{1}{\sqrt{2}}(|0\rangle^{\otimes n}+|1\rangle^{\otimes n})_{ab\ldots n}$ is shared among $N$ parties. This shared state acts as a control qubit to conditionally manipulate the operation order of single-qubit gates by using a quantum switch held by each party. Finally, when the control qubits are measured, the state of the system can be projected into an entangled state on qubits $A$ to $N$.}
\label{SwitchN}
\end{center}
\end{figure}

As shown in Fig. \ref{SwitchN}, the method can be generalized to generate a heralded $N$-qubit entangled state on $N$ separable long-distance qubits by using $N$ quantum switches and a prior shared $n$-qubit Greenberger-Horne-Zeilinger (GHZ) state $|\varphi\rangle_{ab\ldots n}=\frac{1}{\sqrt{2}}(|0\rangle^{\otimes n}+|1\rangle^{\otimes n})_{ab\ldots n}$ among $N$ parties.
Each party possesses qubits $i$ and $I$ ($i=a, \ldots, n$ and $I=A, \ldots, N$), where qubit $i$ acts as a control qubit and qubit $I$ acts as a target qubit. In each quantum switch, when control qubit is in the state $|0\rangle_i$  or $|1\rangle_i$, the different operation orders $U_{I_2}U_{I_1}$ or $U_{I_1}U_{I_2}$  are applied to qubit $I$.
Finally, all control qubits are measured in the basis $\{|\pm\rangle\}$, and then the state of the whole system can collapse into an entangled state by adjusting the coefficients of the initial states and rotation angle $\theta$ of the single-qubit gates $U_{I_1}$.
Here all single-qubit gates are defined as $U_{I_1}=R_y(2\theta)$ and $U_{I_2}=X$.

\section{Optical realization of entangled state generation on separable long-distance qubits} \label{Sec3.1}

\begin{figure} 
\begin{center}
\includegraphics[width=8.7 cm,angle=0]{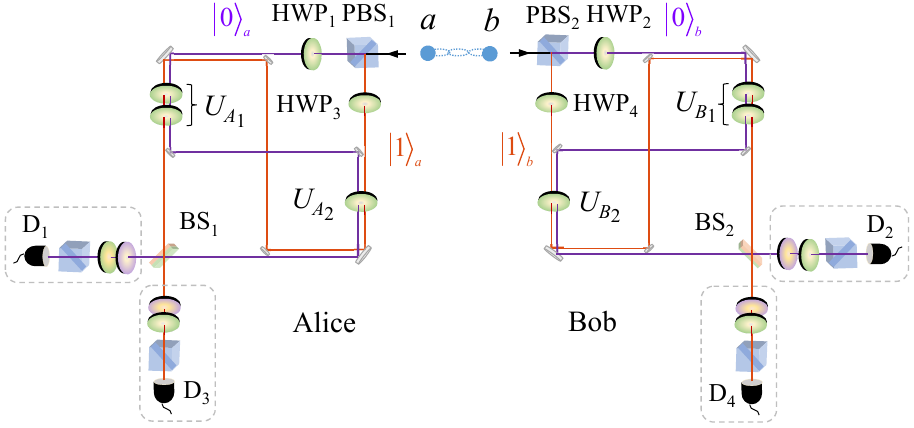}
\caption{Schematic setup to generate optical entangled state on two separable long-distance qubits. A maximally entangled photon pair in the polarization state is shared with Alice and Bob. The paths of the photons are divided into $|0\rangle_a$, $|1\rangle_a$, $|0\rangle_b$, and $|1\rangle_b$ by two polarizing beam splitters, PBS$_1$ and PBS$_2$, because PBS transmits $H$-photon and reflects $V$-photon. The photons emitting from different paths suffer from different gate orders, i.e., $U_{A_2}U_{A_1}$ ($U_{B_2}U_{B_1}$) or $U_{A_1}U_{A_2}$ ($U_{B_1}U_{B_2}$) and then mix at beam splitters BS$_1$ and BS$_2$. Before each detection of path qubits, a polarization analyzer, consisting of a quarter-wave plate (QWP), a half-wave plate (HWP), and a PBS, is inserted to measure the polarization state in an arbitrary basis.}
\label{optics1}
\end{center}
\end{figure}

The generation of entangled states on long-distance qubits can be realized via hybrid encoding in an optical system. Figure \ref{optics1} illustrates an optical architecture for generating an entangled state acting on two long-distance separable polarization qubits.  In this setup, the optical quantum switch is implemented with the path DoF of single photons as the control qubit, and it conditionally manipulates the order of the ``target" single-qubit polarization gates in different propagation directions  \cite{procopio2015experimental,guo2020experimental,rozema2024experimental}.
Let us now present the details of the scheme, step by step.

First, a maximally entangled photon pair in the polarization state
\begin{eqnarray}           \label{eq25}
|\Upsilon_0\rangle=\frac{1}{\sqrt{2}}(|H\rangle_a|H\rangle_b +  |V\rangle_a|V\rangle_b),
\end{eqnarray}
is shared with Alice and Bob. Here $H$ and $V$ denote horizontal and vertical polarization states.
In experiments, the polarization entangled state $|\Upsilon_0\rangle$ can be generated by a pulsed laser pumping into the $\beta-$BaB$_2$O$_4$ (BBO) crystal  \cite{kwiat1999ultrabright,zhang2021spontaneous} with type-I SPDC process or through type-I SPDC in a periodically poled potassium titanyl phosphate (PPKTP) crystal \cite{jin2014pulsed,lee2016polarization,chen2020verification}.
The photons $a$ and $b$ pass through two polarizing beam splitters, PBS$_1$ and PBS$_2$.
PBS transmits the $H$-component and reflects the $V$-component of photon, respectively.
Hence, PBS splits the path of the photon into the path states $|0\rangle_a$, $|1\rangle_a$ $|0\rangle_b$, and $|1\rangle_b$.
After photons pass through PBS$_1$ and PBS$_2$, the state of the polarization-entangled photons can be rewritten as
\begin{eqnarray}           \label{eq26}
|\Upsilon_1\rangle=\frac{1}{\sqrt{2}}(|H\rangle|0\rangle_a|H\rangle|0\rangle_b +  |V\rangle|1\rangle_a|V\rangle|1\rangle_b).
\end{eqnarray}

In order to obtain a general initial state with the same form as the state shown in Eq. \eqref{eq4},
we arrange half-wave plate HWP$_1$ rotated at $\lambda_1^\circ$ in the paths $|0\rangle_a$ and arrange HWP$_3$ rotated at $(\lambda_1+45)^\circ$ in the paths $|1\rangle_a$ to change the $H$-photon into the state $|\phi\rangle_A=\cos2\lambda_1|H\rangle+\sin2\lambda_1|V\rangle$ and change the $V$-photon into $|\phi\rangle_A$.
We arrange HWP$_2$ rotated at $\lambda_2^\circ$ in the paths $|0\rangle_b$ and HWP$_4$ rotated at $(\lambda_2+45)^\circ$ in the paths $|1\rangle_b$ to change the $H$-photon into the state $|\phi\rangle_B=\cos2\lambda_2|H\rangle+\sin2\lambda_2|V\rangle$ and change the $V$-photon into $|\phi\rangle_B$.
These elements (HWP$_1$, HWP$_2$, HWP$_3$, and HWP$_4$) convert $|\Upsilon_1\rangle$  to
\begin{eqnarray}           \label{eq27}
|\Upsilon_2\rangle=\frac{1}{\sqrt{2}}(|0\rangle_a|0\rangle_b +|1\rangle_a|1\rangle_b)\otimes|\phi\rangle_A\otimes|\phi\rangle_B.
\end{eqnarray}
Based on Eq. \eqref{eq25} to Eq. \eqref{eq27}, we can find that the polarized entangled state shown in Eq. \eqref{eq26} is transformed into path entangled state shown in Eq. \eqref{eq27}. Such transformation indicates that the preparation of the initial state is completed.

Second, the path states act as control qubits to manipulate the operation order of polarization gates on states $|\phi\rangle_A$ and $|\phi\rangle_B$.
If the photon is in path $|0\rangle_a$ (or $|0\rangle_b$), it suffers from $U_{A_1}$ (or $U_{B_1}$) followed by $U_{A_2}$ (or $U_{B_2}$).
If the photon is in path $|1\rangle_a$ (or $|1\rangle_b$), it passes through  $U_{A_2}$ (or $U_{B_2}$) followed by $U_{A_1}$ (or $U_{B_1}$).
The operations $U_{A_1}$, $U_{B_1}$, $U_{A_2}$, $U_{B_2}$ are the single-qubit polarization gates.
For $U_{A_2}=U_{B_2}=X$ case, they can be easily realized by an HWP rotated at $45^\circ$.
For $U_{A_1}=U_{B_1}=R_y(2\theta)$ case, they can be realized by an HWP rotated at $0^\circ$ and an HWP rotated at $\frac{\theta}{2}^\circ$.
Then the photons from two paths $|0\rangle_a$ and $|1\rangle_a$ ($|0\rangle_b$ and $|1\rangle_b$) arrive at a balanced beam splitter BS$_1$ (BS$_2$), simultaneously.
These optical elements $U_{A_1}$, $U_{A_2}$, $U_{B_1}$, $U_{B_2}$, BS$_1$, and BS$_2$ transform  $|\Upsilon_2\rangle$ into
\begin{eqnarray}
\begin{split}             \label{eq28}
|\Upsilon_3\rangle=&\frac{1}{2\sqrt{2}}\big[
 |0\rangle_a |0\rangle_b (U_{A_2} U_{A_1} \otimes U_{B_2} U_{B_1}\\& +  U_{A_1} U_{A_2} \otimes U_{B_1} U_{B_2})|\phi\rangle_A \otimes |\phi\rangle_B \\&
+|1\rangle_a |1\rangle_b (U_{A_2} U_{A_1} \otimes U_{B_2} U_{B_1}\\& +  U_{A_1} U_{A_2} \otimes U_{B_1} U_{B_2})|\phi\rangle_A \otimes |\phi\rangle_B \\&
+|0\rangle_a |1\rangle_b (U_{A_2} U_{A_1} \otimes U_{B_2} U_{B_1}\\& -  U_{A_1} U_{A_2} \otimes U_{B_1} U_{B_2})|\phi\rangle_A \otimes |\phi\rangle_B \\&
+|1\rangle_a |0\rangle_b (U_{A_2} U_{A_1} \otimes U_{B_2} U_{B_1}\\& -  U_{A_1} U_{A_2} \otimes U_{B_1} U_{B_2})|\phi\rangle_A \otimes |\phi\rangle_B \big].
\end{split}
\end{eqnarray}
Here BS$_1$ (BS$_2$) performs a Hadamard operation on path state, and the transformation matrix of BS$_1$ (BS$_2$) in the basis $\{|0\rangle_a, |1\rangle_a\}$ ($\{|0\rangle_b, |1\rangle_b\}$) is given by
\begin{eqnarray}
\begin{split}             \label{eq29}
&U_\text{BS}=\frac{1}{\sqrt{2}}\left(\begin{array}{cc}
             1  &   1  \\
             1  &   -1  \\
\end{array}\right).\\
\end{split}
\end{eqnarray}

Finally, Alice and Bob perform a single-qubit path state measurement. The four coincidence clicks of single photon detectors, $D_1$--$D_2$, $D_3$--$D_4$, $D_1$--$D_4$, and $D_3$--$D_2$, correspond to the detection of path states $|0\rangle_a|0\rangle_b$, $|1\rangle_a|1\rangle_b$, $|0\rangle_a|1\rangle_b$, and $|1\rangle_a|0\rangle_b$, respectively. Before each detector, the qubit analyzer consisting of a quarter-wave plate (QWP), an HWP, and a PBS is used to reconstruct the final polarization state in an arbitrary basis.

\begin{figure} 
\begin{center}
\includegraphics[width=7.8 cm,angle=0]{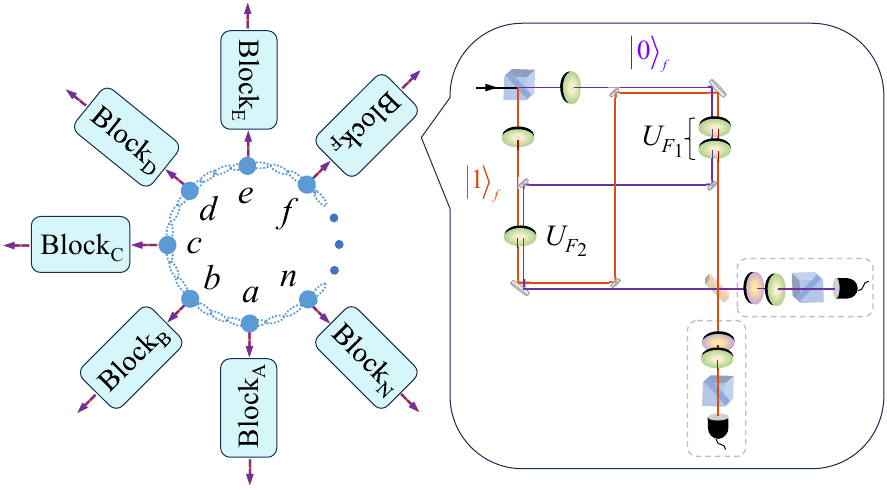}
\caption{Schematic setup to generate optical entangled state on $N$ separable long-distance qubits.}
\label{OpticsN}
\end{center}
\end{figure}

We also devise an optical scheme for generating entangled states on $N$ separable long-distance photonic qubits, as shown in Fig. \ref{OpticsN}.
A  prior $n$-qubit GHZ polarization state $|\Upsilon_0\rangle_{ab\ldots n}=\frac{1}{\sqrt{2}}(|H\rangle^{\otimes n}+|V\rangle^{\otimes n})_{ab\ldots n}$ is shared among $N$ parties.
Each ``Block" transforms the polarization entangled state into the path entangled state $|\Upsilon_1\rangle_{ab\ldots n}=\frac{1}{\sqrt{2}}(|0\rangle^{\otimes n}+|1\rangle^{\otimes n})_{ab\ldots n} \otimes |\phi\rangle_A\otimes |\phi\rangle_B\otimes\ldots\otimes |\phi\rangle_N$,
where the path DoF acts as control qubit to manipulate the order of the single-qubit gates ($U_{I_1}$ and $U_{I_2}$) applied on the polarization state $|\phi\rangle_I$ ($I=A, B, \dots, N$).
Here $|\phi\rangle_I=\cos2\lambda_i|H\rangle+\sin2\lambda_i|V\rangle$ denotes a general single-qubit polarization state. Each single-qubit gate $U_{I_1}=R_y(2\theta)$ can be realized by an HWP rotated at $0^\circ$ and an HWP rotated at $\frac{\theta}{2}^\circ$, and each single-qubit gate $U_{I_2}=X$ can be realized by an HWP rotated at $45^\circ$.
Finally, the path DoFs of the photons are measured, and the $N$-qubit polarization state can be entangled by selecting the appropriate parameters.

\section{Discussion and summary}

\subsection{The performance of the protocols}

Our deterministic protocols in principle with unity fidelity and efficiency. However, in actual  long-distance photon transmission and manipulation, inevitable experimental errors will be introduced and degrade the efficiency of entanglement preparation. The efficiency of the protocol is mainly affected by the transmission loss and scattering loss of photons in the quantum channel, as well as single photon detection. We assume that the distance between each two parties is $L$. In previous protocols, an $N$-qubit entangled state among distant nodes is generated by using a single photon as a common-data bus to couple with $N$ stationary qubits, resulting in a total optical channel length of $(N-1)L$. Using this approach, the efficiency for generating such an $N$-qubit entangled state is given by \cite{xie2023heralded,zhou2023parallel,li2024heralded}
\begin{eqnarray}           \label{eq30}
\eta_N'(L)=\eta_0^N\eta_d\text{exp}[-\alpha(N-1)L].
\end{eqnarray}
Here $\eta_0=0.96$ is the probability of a photon into the optical channel after scattering in photon-cavity interaction. $\eta_d$ is the efficiency of a single-photon detector and $\eta_d\geq0.96$. The exponential term represents photon loss in an optical channel, and $\alpha\simeq \frac{1}{22}$km$^{-1}$ is attenuation rate of photon in the optical channel \cite{sangouard2011quantum}.

In contrast, our protocol employs an $N$-photon state that is shared among the $N$ parties.  By strategically positioning the shared state at a central location relative to all parties, the total transmission distance  of the $N$ photons can be minimized, and the minimum distance is $\frac{N}{2\text{sin}(\frac{\pi}{N})}L$. The scattering loss of a photon in few linear-optics elements in our protocols can be negligible.  Therefore, the efficiency of our protocol for generating an $N$-qubit entangled state among $N$ parties is
\begin{eqnarray}           \label{eq31}
\eta_N(L)=\eta_d^N\text{exp}\bigg[-\alpha\frac{N}{2\text{sin}(\frac{\pi}{N})}L\bigg].
\end{eqnarray}

\begin{figure} 
\begin{center}
\includegraphics[width=6.5 cm,angle=0]{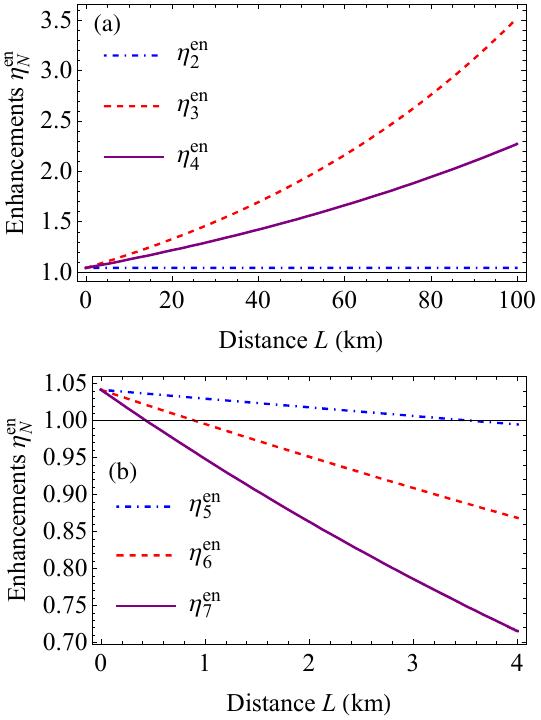}
\caption{The efficiency enhancements $\eta_N^\text{en}$ for generating an $N$-qubit long-distance entangled state among $N$ parties. $L$ represents the distance between each pair of parties. (a) The efficiency enhancements for $N=2, 3, 4$  significantly improve efficiency regardless of the distance $L$. (b) The efficiency enhancements for $N=5, 6, 7$  improve efficiency when the distance $L$ is small.}
\label{enhancement}
\end{center}
\end{figure}

We define an efficiency enhancement $\eta_N^\text{en}(L)=\frac{\eta_N(L)}{\eta_N'(L)}$ to quantify the advantages of our protocols. The enhancement $\eta_N^\text{en}(L)$ is described by
\begin{eqnarray}           \label{eq32}
\eta_N^\text{en}(L)=\eta_0^{-N}\eta_d^{N-1}\text{exp}\bigg[\alpha L \bigg(N-1-\frac{N}{2\text{sin}(\frac{\pi}{N})}\bigg)\bigg].
\end{eqnarray}
The efficiency enhancements as functions of distance $L$ and $N$ are plotted in Fig. \ref{enhancement}. As shown in Fig. \ref{enhancement}(a), when $N=2, 3, 4$, all efficiencies are enhanced regardless of the distance $L$. Specially, $\eta_2^\text{en}(L)=1.042$ for all distances $L$, and $\eta_2'(11)=53.66\%$ and $\eta_2(11)=55.90\%$; $\eta_2'(66)=4.40\%$ and $\eta_2(66)=4.59\%$. When $N=3$ and $N=4$, the efficiencies increase exponentially with the distance $L$ and increases faster for $N=3$. Specially, $\eta_3'(11)=31.25\%$, $\eta_3(11)=37.21\%$, and $\eta_3^\text{en}(11)=1.191$; $\eta_3'(66)=0.21\%$, $\eta_3(66)=0.49\%$, and $\eta_3^\text{en}(66)=2.333$; $\eta_4'(11)=18.19\%$, $\eta_4(11)=20.65\%$, and $\eta_4^\text{en}(11)=1.135$; $\eta_4'(66)=0.01\%$, $\eta_4(66)=0.0175\%$, and $\eta_4^\text{en}(66)=1.748$.
The efficiency enhancements for more qubit entanglement preparation are shown in Fig. \ref{enhancement}(b). When $N=5, 6, 7$, the efficiencies of our protocols can be improved over shorter distances, while the efficiencies will be decreased over longer distances. Specially, when $0<L<3.546$, the efficiency for $N=5$ in our protocol is improved. When $0<L<0.898$, the efficiency for $N=6$ in our protocol is improved. When $0<L<0.435$, the efficiency for $N=7$ in our protocol is improved. Since our protocol involves an $N$-photon state shared among the $N$ parties, it minimizes the total transmission distance of the $N$ photons, resulting in a shorter total distance compared to previous protocols \cite{xie2023heralded,zhou2023parallel,li2024heralded}. This reduction in transmission distance leads to lower photon loss, particularly for small $N$, thereby enhancing the overall preparation efficiency.

\subsection{Potential applications and summary}

Our protocols can generate long-distance MESs, PESs, and SSs, respectively. It is noted that the types of generated states are determined by the coefficients of the initial states and the rotation angles of the single-qubit gates. MESs have been recognized as the essential resources for various quantum information processing tasks. For instance, long-distance maximally entangled Bell states and GHZ states are crucial for implementing quantum gates \cite{zeuner2018integrated}, quantum secure direct communication \cite{jin2006three,zhou2020device}, and quantum cryptography \cite{yin2020entanglement}. Furthermore, PESs, such as less-entangled Bell states and GHZ-like states, play a significant role in remote quantum state preparation \cite{hua2016deterministic} and multi-party quantum communication \cite{dong2011controlled}.

Remarkably, our protocols employing the ICO approach generate states that can be equivalent to the graph states up to local single-qubit unitary transformations. In fact, an $n$-qubit graph state $|G\rangle=(V,E)$ is defined as \cite{hein2004multiparty}
\begin{eqnarray}           \label{eq45}
|G\rangle=\prod_{(i,j)\in E}\text{CZ}^{(i,j)}|+\rangle^{\otimes n},
\end{eqnarray}
where each vertex ($V$) represents one qubit and each edge ($E$) corresponds to a pairwise entangling operation. CZ$^{(i,j)}$ denotes a controlled phase-flip gate acting on qubits $i$ and $j$. In our protocols, if the pre-shared Bell state $|\varphi\rangle_{ab}$ is replaced by the entangled state $|\varphi'\rangle_{ab}=\frac{1}{\sqrt{2}}(|0\rangle_{a}|0\rangle_{b}+\texttt{i}|1\rangle_{a}|1\rangle_{b})$,
and after applying the same processes described in Sec. \ref{Sec2.1}, an ICO operation on $|\psi\rangle_A|\psi\rangle_B$ becomes
\begin{eqnarray}
\begin{split}        \label{eq46}
\text{ICO}^{(1,2)}_\pm=&\;\frac{1}{\sqrt{2}}[(U_{A_2}U_{A_1})\otimes(U_{B_2}U_{B_1}) \\&
                                     \pm \texttt{i}(U_{A_1}U_{A_2})\otimes(U_{B_1}U_{B_2})].
\end{split}
\end{eqnarray}
Similarly, when all $n$ initial states are chosen as $|\psi\rangle_I=|+\rangle$ ($I=A, B, \ldots, N$) and the pre-shared state among the $N$ parties is $|\varphi'\rangle_{ab\ldots n}=\frac{1}{\sqrt{2}}(|0\rangle^{\otimes n}+\texttt{i}|1\rangle^{\otimes n})_{ab\ldots n}$, then, after applying the same ICO operations and measuring the shared qubits  in the $\{|\pm\rangle\}$ basis, our protocols can generate the state
\begin{eqnarray}
\begin{split}        \label{eq47}
|G'\rangle=\prod_{(i,j)\in E}\text{ICO}_\pm^{(i,j)}|+\rangle^{\otimes n}.
\end{split}
\end{eqnarray}
It is straightforward to verify that when $U_{A_i}=U_{B_i}=R_z(\pi/2)$ and $U_{A_j}=U_{B_j}=X$, the operations $\text{CZ}^{(i,j)}$ and $\text{ICO}^{(i,j)}_\pm$ satisfy the following identity \cite{liu2025quantum}
\begin{equation}  \label{48}
\text{CZ}^{(i,j)}=\left\{\begin{aligned}
&e^{-\frac{\pi}{4}\texttt{i}}(\sigma_y\otimes \sigma_y)\text{ICO}^{(i,j)}_{+},  \\
&e^{\frac{\pi}{4}\texttt{i}}(\sigma_x\otimes \sigma_x)\text{ICO}^{(i,j)}_{-}. \\
\end{aligned}
\right.
\end{equation}
Therefore, the state $|G'\rangle$ generated by our protocols is equivalent to the graph state $|G\rangle$ up to local unitary transformations.

Our protocols allow for the entanglement of $n$ independent target quantum states over long distances among multiple parties via shared entanglement (see Fig. \ref{SwitchN} and Fig. \ref{OpticsN}), and they provide an alternative approach for constructing large-scale quantum networks. This approach has some important advantages: (1) The initial states at different nodes do not require direct interaction, instead they rely on the superposition of the orders of single-qubit gates. This feature makes our approach particularly well-suited for establishing large-scale optical quantum networks, because direct photon-photon interactions are difficult to achieve. (2) The method exhibits excellent scalability and operations at each node are performed independently, which effectively reduces qubit crosstalk and enhances the overall robustness of the system. (3) As described above, our approach can generate arbitrary distributed graph states within quantum networks \cite{meignant2019distributing,koudia2023deterministic,lin2025enhanced}. These are a critical resource for measurement-based quantum computation and also are a promising application for implementing graph-state-based communication protocols \cite{markham2008graph,bell2014experimental}.

In summary, we have presented protocols for deterministically generating an $N$-qubit entanglement among $N$ remote parties. A pre-shared maximally entangled state is used to control the orders of single-qubit gates on each target qubit at each party with ICO.  Unlike the entanglement swapping technique, our approach avoids multiple pre-shared entanglements and complex entangled state measurements, relying instead on one pre-shared entanglement and simple single-qubit gates, which enhances experimental feasibility and flexibility. Interestingly, this approach generates state that can be equivalent to graph state, and extending the framework to generate hypergraph state \cite{huang2024demonstration} remains an interesting avenue for future research.

In addition, we propose compact optical schemes for implementing these protocols by encoding information in the polarization and path DoFs of photons. Unlike previous approaches based on photon-spin interactions \cite{xie2023heralded,zhou2023parallel,li2024heralded}, the optical schemes rely on a pre-shared $N$-photon state and the superposition of single-qubit gate operations, which offers a fundamentally different approach. This key distinction leads to higher efficiency and simplified state preparation. For example, the efficiencies of our schemes for generating 3-qubit and 4-qubit entanglements are improved exponentially with the distance $L$.  For larger numbers of qubits, efficiencies also can be improved over the short distances. This work presents an efficient and practical framework for general entanglement generation among multiple nodes, serving as an essential foundation for the realization of large-scale quantum networks. This advancement paves the way for the development of practical quantum technologies, with potential applications in distributed quantum computing, remote state preparation,
and multi-party long-distance quantum communication.

\section*{ACKNOWLEDGMENTS}

This work was funded by Science Research Project of Hebei Education Department under Grant No. QN2025054, and National Natural Science Foundation of China under Grant No. 62371038.

%


\end{document}